\documentclass[showpacs,preprintnumbers,amsmath,amssymb,epsfig]{revtex4}
\usepackage{graphicx}% Include figure files
\usepackage{dcolumn}% Align table columns on decimal point
\usepackage{bm}% bold math

\begin{document}
\title{Generation of two-mode nonclassical states and a quantum
phase gate operation in trapped ion cavity QED}
\author{XuBo Zou, K. Pahlke and W. Mathis }
\address{Electromagnetic Theory Group at THT,\\
 Department of Electrical
Engineering, University of Hannover, Germany}

\date{\today}

\begin{abstract}
We propose a scheme to generate nonclassical states of a quantum
system, which is composed of the one-dimensional trapped ion
motion and a single cavity field mode. We show that two-mode SU(2)
Schr\"odinger-cat states, entangled coherent states, two-mode
squeezed vacuum states and their superposition can be generated.
If the vibration mode and the cavity mode are used to represent
separately a qubit, a quantum phase gate can be implemented.
%\vspace{0.25cm}
%PACS number(s):
\end{abstract}
\maketitle

In recent years, there is much interest in methods of nonclassical
state generation by entangling quantum systems. In this field
advances in ion cooling and ion trapping opened new prospects by
controlling the quantized ion motion precisely
\cite{MM}\cite{CCVPZ}\cite{Gerry2}. In particular, motional Fock
states, squeezed states and Schr\"{o}dinger-cat states are in the
scope. More recently, various schemes were proposed to control two
vibration-modes of a single ion, which is confined in a
two-dimensional trap \cite{Gerry2}\cite{GN}\cite{GGM}\cite{zou}.
Most of these proposals concern the quantized ion vibration by
treating the laser field classicaly. But, the quantization of the
laser field brings more possibilities. A number of schemes were
proposed for generating various nonclassical states of single mode
cavity fields \cite{a1}. An experimental realization of a
Schr\"odinger-cat state in a cavity system was reported \cite{a2}.
Furthermore, a single trapped ion can be used as a probe for
measuring the cavity field intensity \cite{walter}. A new
possibility for quantum state engineering and quantum information
processing is opened by using a trapped ion in a high-Q cavity.
The influence of the field statics on the ion dynamics \cite{zz}
was investigated as well as the transfer of the coherence between
the motional state and the light field \cite{pp}. In a recent
paper \cite{ss} a simple scheme was proposed to generate the whole
Bell basis composed of the vibrational mode of a single ion and a
cavity field.\\
In this paper, we propose a scheme to generate two-mode
nonclassical states of a quantum system, which is composed of the
vibration mode of one trapped ion and a single light field mode.
The set of these states includes two-mode SU(2) Schr\"odinger-cat
states \cite{Sanders2}\cite{GGJMO}, entangled coherent states
\cite{Sanders}\cite{Chai}, two-mode squeezed vacuum states and
their superpositions \cite{GGJMO}. We also show that quantum phase
gates can be implemented by representing the qubits by the
vibration mode of a single trapped ion and the cavity mode quantum
state restricted on the subspace spanned by the two lowest Fock
states. In order to relate our scheme to this physical
arrangement, we consider a trapped ion confined in a harmonic trap
located inside an optical cavity. The atomic transition between
the two internal electronic states $|e\rangle$ and $|g\rangle$
(frequency $\omega_0$) is coupled to a single mode of the cavity
field of the frequency $\omega_{cav}$ and is also driven by an
external classical laser field of the frequency $\omega_A$. The
cavity is aligned along the $x$-axis, while the laser field is
incident from a direction along the $y$-axis. Thus, in a frame
rotating at the laser frequency $\omega_A$ the system's
Hamiltonian is in the form \cite{pp}
\begin{eqnarray}
H&=&\nu{b^{\dagger}b}+\delta_{cA}{a^{\dagger}a}+\Delta_{oA}\sigma_+
\sigma_-+E_A\sigma_++E_A\sigma_-\nonumber\\
&&+g_0\sin\eta({b^{\dagger}+b})({a^{\dagger}\sigma_-+a\sigma_+})
\,.\label{1}
\end{eqnarray}
Here $(a,a^{\dagger})$ and $(b,b^{\dagger})$ are the boson
annihilation and creation operators of the cavity field and the
quantized atomic vibration. The operator
$\sigma_-=|g\rangle\langle e|$ changes the internal electronic
state from $|e\rangle$ to $|g\rangle$ and $\eta$ is the
corresponding Lamb-Dicke parameter. The detuning $\delta_{cA}$ and
$\Delta_{oA}$ are given by $\delta_{cA}=\omega_{cav}-\omega_{A}$
and $\Delta_{oA}=\omega_{o}-\omega_{A}$. The quantity $E_A$ is the
amplitude of the laser field. The single photon atom-cavity dipole
coupling strength is given by $g_0$, while the $sin$-function
describes the standing wave structure of the cavity field. We
assume that the center of the trap is located at the node of the
cavity field. Since the detuning of the laser field from the
atomic transition frequency is assumed to be very large
($\Delta_{oA}>>\nu,\delta_{cA},g_0, E_A$), the internal atomic
dynamics can be adiabatically eliminated. The corresponding
Hamiltonian takes the form
\begin{eqnarray}
H&=&\nu{b^{\dagger}b}+\delta_{cA}a^{\dagger}a
-\frac{g_0^2}{\Delta_{oA}}\sin^2\eta(b^{\dagger}+b)a^{\dagger}a\sigma_z\nonumber\\
&&-\frac{g_0\epsilon_A}{\Delta_{oA}}\sin\eta(b^{\dagger}+b)(e^{-i\varphi_A}a^{\dagger}+e^{i\varphi_A}a)\sigma_z
\,,\label{2}
\end{eqnarray} with $\sigma_z=|e\rangle\langle
e|-|g\rangle\langle g|$ and $E_A= \epsilon_Ae^{-i\varphi_A}$. In
the following we choose $\varphi_A=\frac{\pi}{2}$. In the
Lamb-Dicke region ($\eta<<1$) we may write
$\sin\eta(b^{\dagger}+b)\simeq\eta(b^{\dagger}+b)$. If we choose
the detuning between the cavity and the laser field to be
$\delta_{cA}=\nu$ the effective interaction Hamiltonian
\begin{eqnarray}
H_1=i\Omega_1(a^{\dagger}b-ab^{\dagger})\sigma_z \label{3}
\end{eqnarray}
is obtained in the rotating wave approximation. Here $\Omega_1$
denotes the effective interaction strength. We consider the
situation in which the ion's internal state is prepared initially
as a superposition $\frac{1}{\sqrt{2}}(|g\rangle+|e\rangle)$ and
the vibration mode of the ion is in the Fock state $|n\rangle_b$.
The cavity field is prepared in the vacuum state $|0\rangle_a$:
\begin{eqnarray}
\Psi(0)=\frac{1}{\sqrt{2}}(|g\rangle+|e\rangle)\,|0\rangle_a|n\rangle_b\,.
\label{4}
\end{eqnarray}
If the laser pulse of the interaction (3) is applied, the system
evolves into
\begin{eqnarray}
\Psi(t)&=&\frac{1}{\sqrt{2}}\exp[\Omega_1t(a^{\dagger}b-ab^{\dagger})]
\quad |e\rangle|0\rangle_a|n\rangle_b \nonumber\\
&&+\frac{1}{\sqrt{2}}\exp[-\Omega_1t(a^{\dagger}b-ab^{\dagger})]
\quad |g\rangle|0\rangle_a|n\rangle_b\,. \label{5}
\end{eqnarray}
The carrier transition $|e\rangle\leftrightarrow|g\rangle$ is
driven in order to generate a $\frac{\pi}{2}$-pulse on the
internal state:
$|e\rangle\rightarrow\frac{1}{\sqrt{2}}(|g\rangle+|e\rangle)$;
$|g\rangle\rightarrow\frac{1}{\sqrt{2}}(|g\rangle-|e\rangle)$. The
quantum state evolves into
\begin{eqnarray}
\Psi(t)&=&\frac{1}{2}(\exp[\Omega_1t(a^{\dagger}b-ab^{\dagger})]+\exp[-\Omega_1t(a^{\dagger}b-ab^{\dagger})])
\quad|g\rangle|0\rangle_a|n\rangle_b \nonumber\\
&&+\frac{1}{2}
(\exp[\Omega_1t(a^{\dagger}b-ab^{\dagger})]-\exp[-\Omega_1t(a^{\dagger}b-ab^{\dagger})])
\quad|e\rangle|0\rangle_a|n\rangle_b\,. \label{6}
\end{eqnarray}
Upon a detection of the internal state, this state vector is
projected into the two-mode quantum states, which we denote with
\begin{eqnarray}
\Phi_{\pm}=\exp[-i\Omega_1t(a^{\dagger}b-ab^{\dagger})]
\,|0\rangle_a|n\rangle_b\pm
\exp[i\Omega_1t(a^{\dagger}b-ab^{\dagger})]\,
|0\rangle_a|n\rangle_b\,. \label{7}
\end{eqnarray}
The quantum state $\Phi_+$ is obtained if the internal state
$|g\rangle$ of the ion is detected. Otherwise the quantum state
$\Phi_-$ is generated. These two-mode quantum states (7) are the
SU(2)-Schr\"{o}dinger-cat states of the form $(|\zeta,j\rangle \pm
|-\zeta,j\rangle)$ \cite{GGJMO}, if the SU(2)-coherent states
$|\zeta,j\rangle= \exp[\beta J_+ -\beta^* J_- ]|j,-j\rangle$
\cite{Buzek} are used with the identification: $\beta=\Omega_1t$;
$j=\frac{n}{2}$; $\zeta=
\tan{\frac{\Omega_1t}{2}}$.\\

In order to entangle coherent states of a special form, we again
initialize the internal ion state in
$\frac{1}{\sqrt{2}}(|g\rangle+|e\rangle)$ but we prepare the two
bose-modes in the coherent states $|\alpha\rangle_a$ and
$|\beta\rangle_b$:
\begin{eqnarray}
\Psi(0)=\frac{1}{\sqrt{2}}(|g\rangle+|e\rangle)\,|\alpha\rangle_a|\beta\rangle_b
\,.\label{8}
\end{eqnarray}
If the laser pulse, which corresponds to the interaction (3), is
applied over a time interval $t_1=\frac{\pi}{4\Omega_1}$, the
system evolves into
\begin{eqnarray}
\Psi(t_1)&=&\frac{1}{\sqrt{2}}|e\rangle\,|(\alpha+\beta)/\sqrt{2}\rangle_a\,|(\beta-\alpha)/\sqrt{2}\rangle_b
\nonumber\\
&&+\frac{1}{\sqrt{2}}|g\rangle\,|(\alpha-\beta)/\sqrt{2}\rangle_a\,|(\alpha+\beta)/\sqrt{2}\rangle_b
\,.\label{9}
\end{eqnarray}
Then the atom is subjected to a $\frac{\pi}{2}$-pulse, which is
resonant with the transition $|e\rangle\leftrightarrow|g\rangle$.
Upon a detection of the internal state, the state vector is
projected into the two-mode coherent states
\begin{eqnarray}
\Phi_{\pm}&=&|(\alpha+\beta)/\sqrt{2}\rangle_a\,|(\beta-\alpha)/\sqrt{2}\rangle_b\nonumber\\
&&\pm|(\alpha-\beta)/\sqrt{2}\rangle_a\,|(\alpha+\beta)/\sqrt{2}\rangle_b
\,.\label{10}
\end{eqnarray}
This equation demonstrates the generation of entangled coherent
states of a special type. Entangled coherent states of another
type can be generated, if the interaction time is chosen two times
longer: $\Omega_1t_1=\frac{\pi}{2}$. In this example the initial
state (8) evolves into
\begin{eqnarray}
\Psi(t_1)&=&\frac{1}{\sqrt{2}}|e\rangle|\beta\rangle_a|-\alpha\rangle_b\nonumber\\
&&+|g\rangle|-\beta\rangle_a|\alpha\rangle_b\,. \label{11}
\end{eqnarray}
If the atom's internal state is subjected to a
$\frac{\pi}{2}$-pulse, like it is described above, the detection
of the ground state $|g\rangle$ or the excited state $|e\rangle$
projects into the entangled coherent states
\begin{eqnarray}
\Phi_{\pm}=|\beta\rangle_a|-\alpha\rangle_b\pm|-\beta\rangle_a|\alpha\rangle_b
\,.\label{12}
\end{eqnarray}
There is also the possibility of choosing the detuning between
cavity field and laser field to be $\delta_{cA}=-\nu$. In this
case, we can obtain the corresponding effective interaction
Hamiltonian
\begin{eqnarray}
H_2=i\Omega_2(ab-a^{\dagger}b^{\dagger})\sigma_z \,.\label{13}
\end{eqnarray}
Here, $\Omega_2$ denotes the effective interaction strength. The
interaction (12) is a parametric amplifier, which leads to a
two-mode squeezed vacuum state, if the two bose-modes are at the
beginning in the vacuum state and the ion's vibration ground
state. Again we initialize the ion's internal state in the state
$\frac{1}{\sqrt{2}}(|g\rangle+|e\rangle)$. Thus, the system's
initial state
\begin{eqnarray}
\Psi(0)=\frac{1}{\sqrt{2}}(|g\rangle+|e\rangle)\,|0\rangle_a|0\rangle_b
\,.\label{14}
\end{eqnarray}
evolves under the interaction (13) into
\begin{eqnarray}
\Psi(t)&=&\frac{1}{\sqrt{2}}\exp[\Omega_2t(a^{\dagger}b^{\dagger}-ab)]
\,|e\rangle|0\rangle_a|0\rangle_b\nonumber\\
&&+\frac{1}{\sqrt{2}}\exp[-\Omega_2t(a^{\dagger}b^{\dagger}-ab)]
\,|g\rangle|0\rangle_a|0\rangle_b\,. \label{15}
\end{eqnarray}
After subjecting the atom's internal state to a
$\frac{\pi}{2}$-pulse and detecting the internal state, a
projection into the squeezed cat states results
\begin{eqnarray}
\Phi_{\pm}=\exp[\Omega_2t(a^{\dagger}b^{\dagger}-ab)]
|0\rangle_a|0\rangle_b\pm
\exp[-\Omega_2t(a^{\dagger}b^{\dagger}-ab)]|0\rangle_a|0\rangle_b
\,.\label{16}
\end{eqnarray}
Finally we consider the case $\omega_0-\omega_{cav}=\nu$. In the
Lamb-Dicke limit we obtain after discarding the rapidly
oscillating terms in Hamlitonian (1) the interaction Hamiltonian
\cite{ss}
\begin{eqnarray}
H_3=\Omega_3(ab\sigma_++a^{\dagger}b^{\dagger}\sigma_-)\,.
\label{17}
\end{eqnarray}
This was used to generate Bell-states of a single ion vibration
mode and the cavity field state \cite{ss}. When the ion's internal
state is initially in the ground state $|g\rangle$, the
interaction (17) is applied with $\Omega_3t_1=\pi$. We obtain
\begin{eqnarray}
|0\rangle_a|0\rangle_b|g\rangle&\longrightarrow&|0\rangle_a|0\rangle_b|g\rangle
\nonumber\\
|0\rangle_a|1\rangle_b|g\rangle&\longrightarrow&|0\rangle_a|1\rangle_b|g\rangle
\nonumber\\
|1\rangle_a|0\rangle_b|g\rangle&\longrightarrow&|1\rangle_a|0\rangle_b|g\rangle
\nonumber\\
|1\rangle_a|1\rangle_b|g\rangle&\longrightarrow&-|1\rangle_a|1\rangle_b|g\rangle
\,,\label{18}
\end{eqnarray}
where the states to the left of the arrows represent initial
states $\Psi(0)$ and the states to the right are the states
$\Psi(t_1)$. The Eq.(18) shows that a quantum phase gate operation
can be implemented, if the vibration mode and the cavity mode are
used to represent separately a qubit.\\

In summary, we presented a scheme to generate nonclassical
two-mode states of a single trapped ion vibration mode and the
light field state. This quantum state generation scheme makes the
generation of two-mode SU(2)-Schr\"odinger-cat states, entangled
coherent states, two mode squeezed vacuum states and their
superposition possible. We also showed that a quantum phase gate
can be implemented, if the two-qubit states are identified with
the
two-mode Fock states with the quantum numbers $0$ and $1$.\\
These results demonstrate the usefulness of this trapped ion
cavity system in respect of the nonclassical state generation.\\
Note add: After we submitted this manuscript, we noticed two
different proposals to implement quantum logic gates in a trapped
ion cavity QED system \cite{mmm}.

\end{document}